\documentclass{iopart}

\usepackage[english]{babel}
\usepackage{graphicx}
\usepackage{psfrag}
\usepackage{amssymb}
\usepackage{amscd}
\usepackage{eucal}
\usepackage{color}
\usepackage{bm}
\usepackage{lipsum}
\usepackage{hyperref}
\usepackage{todonotes}
\usepackage{iopams}
\usepackage{bm}
\usepackage[ruled,vlined]{algorithm2e}

\begin{document}

\title[Locating line and node disturbances]{Locating line and node disturbances in networks of diffusively coupled dynamical agents}
\author{Robin Delabays\textsuperscript{1,2}, Laurent Pagnier\textsuperscript{3}, and Melvyn Tyloo\textsuperscript{4,5}}
\address{\textsuperscript{1} Automatic Control Laboratory, ETH Zurich, CH-8092 Zurich, Switzerland. \\
\textsuperscript{2} Center for Control, Dynamical Systems and Computation, UCSB, Santa Barbara, CA 93106-5070 USA. \\
\textsuperscript{3} Department of Mathematics, University of Arizona, Tucson, AZ 85721, United States. \\
\textsuperscript{4} Department of Quantum Matter Physics, University of Geneva, CH-1211 Geneva, Switzerland.\\
\textsuperscript{5} School of Engineering, University of Applied Sciences of Western Switzerland HES-SO, CH-1951 Sion, Switzerland.}
\ead{robindelabays@ucsb.edu, laurentpagnier@math.arizona.edu, melvyn.tyloo@gmail.com}

\begin{abstract}
 A wide variety of natural and human-made systems consist of a large set of dynamical units coupled into a complex structure. 
 Breakdown of such systems can have a dramatic impact, as in the case of neurons in the brain or lines in an electric grid, to name but a few. 
 Preventing such catastrophic events requires in particular to be able to detect and locate the source of disturbances as fast as possible. 
 We propose a simple method to identify and locate disturbances in networks of coupled dynamical agents, relying only on time series measurements and on the knowledge of the (possibly Kron-reduced) network structure. 
 The strength and the appeal of the present approach lies in its simplicity paired with the ability to precisely locate disturbances and even to differentiate between line and node disturbances. 
 If we have access to measurement at only a subset of nodes, our method is still able to identify the location of the disturbance if the disturbed nodes are measured. 
 If not, we manage to identify the region of the network where the disturbance occurs.
\end{abstract}

\noindent{\it Keywords\/}: Line disturbance, node disturbance, detection and localization.

\section{Introduction}
Despite recent advances in network science~\cite{strogatz2001exploring,rodrigues2016kuramoto}, the understanding of large, networked dynamical system is still incomplete, even though such systems can play a major role in our daily life. 
Large scale events therein should be monitored or mitigated, if not completely avoided. 
Preventing catastrophic events in a networked system requires, in particular, an accurate and reliable assessment of its state in real time. 
Indeed, the early detection of a source of disturbances and its precise location allow preventive measures to be taken. 

In networks, the most fundamental disturbances can affect either of their two basic constituents, namely either nodes or lines. 
Nodal perturbations, occurring mostly as variations of intrinsic velocities of the disturbed agents, act as an additive perturbations, whereas disturbances on lines, which are modifications of the coupling parameters, are represented as multiplicative disturbances in the model, which are much harder to tackle analytically. 
This partly explains the scarcity of the literature covering line disturbances. 
Nevertheless, such perturbations are at least as important as nodal perturbations and, in our opinion, should be investigated in details. 

Systems that can be modeled as a set of diffusively coupled dynamical agents where state estimation is needed for security, comfort, or diagnostic reasons are extremely diverse. 
To name but a few, they range from neurons in the brain to transportation networks and disease spreading models~\cite{Cas09,Bar16,Str17}. 
The case of high voltage electrical grids is an important example of such systems and has attracted a lot of work along the last years, especially in the context of the ongoing energy transition in many parts of the world~\cite{brummitt2013transdisciplinary,markard2018the}. 
We will systematically refer to the power grid and often use its terminology in this manuscript for illustrative purposes, even though our considerations extend to a much larger class of systems. 

In the context of power grids, recent applications of results from network science and dynamical systems theory allowed the development of new techniques aiming at a fast assessment of the system's state.
Such improvements are based on the measurement of some quantities such as voltage amplitudes, phases, and frequencies, which are nowadays widely accessible on a high time resolution thank, for instance, to Phasor Measurement Units (PMUs).
Part of the work on state estimation, such as Refs.~\cite{dosiek2013mode,lokhov2018online,furutani2019network}, is mostly focused on the estimation of the current operating state of the whole system. 
This is performed using various techniques, such as estimation of the eigenmodes of the network through probing signals~\cite{dosiek2013mode} or resonance methods~\cite{furutani2019network}, or based on the measurement of the response of the system to ambient noise~\cite{lokhov2018online}. 
Closer to our interest in this manuscript, Refs.~\cite{nudell2013a,upadhyaya2015power,mathew2016pmu,lee2018data,semerow2016disturbance} proposed some approaches aiming at locating the source of a disturbance. 
For networks composed of areas with weak inter-area connections and strong intra-area connections, Ref.~\cite{nudell2013a} uses the residues of an estimated transfer function in order to locate a nodal disturbance. 
Other approaches rely on \emph{Discrete Wavelet Transform (DWT)}~\cite{upadhyaya2015power,mathew2016pmu} or logistic regression~\cite{lee2018data} of PMU measurements in order to identify the source of a disturbance. 
Most of these methods are designed to locate a nodal disturbance only.

Despite their relevance in electrical networks~\cite{chakravarthy1998nonlinear,horak2004a,hassan2011review}, the case of line disturbances led to fewer results compared to nodal disturbances~\cite{coletta2018performance,delabays2019rate}, and less focus was put on assessing the impact of a line perturbation.  
Methods based on the disturbance propagation~\cite{semerow2016disturbance} would be able to determine the area where a disturbed line is located, but to the best of our knowledge, there are no such methods specifically designed to locate line disturbances. 

Recent works analyzed the case of line failures in power grids, e.g., following an adversary attack~\cite{brahma2011fault,soltan2017analyzing,soltan2018power,soltan2019expose,soltan2019react,jamei2018low-resolution,jamei2020phasor}.
These approaches assumes full knowledge of the system prior to the line failure, and analyzes the mismatch between the expected and actual voltages and powers to recover the set of failed lines, by optimizing over the potential fault locations. 
On top of the line failures, the authors of~\cite{soltan2017analyzing,soltan2018power,soltan2019expose,soltan2019react} assume that the attacker is able to block data acquisition in the area of the fault, and sometimes even sends some spurious data to the monitoring device. 
In the approach of~\cite{jamei2018low-resolution,jamei2020phasor}, the authors assume only partial access to measurements, and find the most likely location for the failure. 
However, in these approaches, the authors assume that the system operator is aware of the attack, and their analysis follows the line failure, which means that dramatic events already happened. 

In this manuscript, we propose to detect and locate disturbances whose impact on the system is not (yet) threatening its operation, without any a priori knowledge of the faults' characteristics or approximate location. 
Our method relies on the measurements of the state of the dynamical agents composing the system and on the knowledge of the interactions between them when the system is unperturbed. 
This means that, in theory, our method can be applied online, without requiring to know if the system is perturbed or not. 
Conveniently, we cover the case of disturbed nodes as well as disturbed lines in a unified framework. 
Assuming that the amplitude and the rate of change of these disturbances are not too large, we propose a way to locate the faulty elements (nodes and/or lines) in the network, based on measurement of the nodes' trajectories. 
In a similar spirit as Refs.~\cite{soltan2017analyzing,soltan2018power,soltan2019expose,soltan2019react}, but in a more general framework, we locate the source of the disturbances by inspection of the mismatch between expected and actual agents' velocities. 
If we have access to measurements at the faulty elements, we are able to precisely locate it, even though such slow disturbance will spread throughout the whole network and have a similar impact on each agent. 
In case of partial measurements, we are nevertheless able to determine the area of the network where the faulty element is located. 
Furthermore, our implementation is adapted to a continuous monitoring of the system and can be used to preventively detect and locate deficient elements before the occurrence of a more significant event, such as a cascading failure, and that at a rather low cost as our method does not rely on any optimization but only on a single matrix-vector product at each time step.

The manuscript is organized as follows. 
We recall some preliminary tools and give a description of the type of models considered in Section~\ref{sec:prelim}. 
In Sections~\ref{sec:method} and \ref{sec:multi_dist}, we detail the method to locate the faulty element for a single and multiple disturbances, and numerical illustrations are given in Section~\ref{sec:nums}.

\section{Notations and framework}\label{sec:prelim}
Throughout this manuscript, we denote by $\bm{e}_i$ the $i$th vector of the canonical basis, with $1$ at index $i$ and zero everywhere else, and $\bm{e}_{ij}=\bm{e}_i-\bm{e}_j$. 
We write ${\rm diag}(\{m_i\})\in \mathbb{R}^{n\times n}$ for the diagonal matrix with elements $m_1, m_2,..., m_n$ on its diagonal and $\bm{1}_n$ (resp. $\bm{0}_n$) the vector of ones (resp. zeros) of length $n$, where the subscript is omited when the dimension is clear from the context. 
Finally, we denote by $\mathbb{I}$ the identity matrix. 

\subsection{Networked diffusive systems}\label{sec:model}
We model the dynamics of a set of $n\in\mathbb{N}$ diffusively interacting agents as 
\begin{eqnarray}\label{eq:dyn}
 m_i\ddot{x}_i + d_i\dot{x}_i = \omega_i - \sum_ja_{ij}f(x_i - x_j)\, , \qquad i \in \{1,...,n\}\, ,
\end{eqnarray}
where the real numbers $m_i$, $d_i$, and $\omega_i$ are respectively the inertia, the damping, and the natural velocity of agent $i$, $a_{ij}\in\mathbb{R}$ is the $(i,j)$th element of the weighted adjacency matrix of the interaction graph, and $f\colon\mathbb{R}\to\mathbb{R}$ is an odd, differentiable coupling function, such that $f'(0)>0$. 
We assume the interaction graph to be undirected, i.e., $a_{ij}=a_{ji}$ (the directed case is discussed in the Appendix). 
Note that we will be interested in fixed points of Eq.~(\ref{eq:dyn}), i.e., when the left hand side vanishes, meaning that all our results apply to models with higher order time derivatives. 

Assume Eq.~(\ref{eq:dyn}) has a fixed point, denoted $\bm{x}^*\in\mathbb{R}^n$. 
When the system is not too heterogeneous, i.e., the spread of $\{\omega_i\}$ is small relatively to the magnitude of the couplings, the fixed point components are small (in absolute value) and are well-approximated by the linearized equation 
\begin{eqnarray}\label{eq:lin_fix}
 \bm{\omega} = \bm{L} \bm{x}\, ,
\end{eqnarray}
where $\bm{L}\in\mathbb{R}^{n\times n}$ is the Jacobian matrix defined as,
\begin{eqnarray}
L_{ij} = -\frac{\partial F_i(\bm{0})}{\partial x_j}\,, \qquad i,j\in\{1,...,n\}\, ,
\end{eqnarray}
where $F_i(\bm{x})$ is given by the sum in the right-hand side of Eq.~(\ref{eq:dyn}). 
This Jacobian matrix satisfies the properties (i) $\sum_j L_{ij}=0$ and (ii) $L_{ii}=-\sum_j L_{ij}$ $\forall i$, it is then a (weighted) Laplacian matrix corresponding to the interaction graph of the system Eq.~(\ref{eq:dyn}), with weights related to the strength of the couplings between agents. 
This approximation is common in the context of power grids, where it is called \emph{DC power flow equations}~\cite[Sec.~9.7]{grainger1994power}. 
Obviously, for linear coupling functions, Eq.~(\ref{eq:lin_fix}) is not an approximation. 

{\bf Remark.} 
{\it Throughout this manuscript, we will refer to dynamical agents as \emph{nodes} and to the connections between them as \emph{lines}. 
In the context of graph theory, these two objects are usually called \emph{vertices} and \emph{edges} respectively. }

\subsection{Disturbances}\label{ssec:dist}
We consider disturbances either at nodes, yielding the time varying natural velocity 
\begin{eqnarray}
 \omega_i(t) = \omega_i^0 + \xi_{\rm n}(t)\, ,
\end{eqnarray}
or on lines, described as varying coupling 
\begin{eqnarray}
 a_{ij}(t) = a_{ij}^0 + \xi_{\rm l}(t)\, ,
\end{eqnarray}
where we take $\xi_{\rm n,l}(t)$ to be any function of time, whose rate of change is typically smaller than any intrinsic time scale of the system Eq.~(\ref{eq:dyn}). 
Formally, we require that the disturbance's rate of change is smaller than the rate at which its effect is damped at the agents and smaller than the speed at which it spreads throughout the network, i.e., for all $i,j$, 
\begin{eqnarray}\label{eq:ass_freq}
 \max_t|\dot{\xi}_{\rm n,l}(t)| &\ll \min\left\{\frac{d_i}{m_i}, \frac{\lambda_j}{\sqrt{m_i}}, \frac{\lambda_j}{d_i}\right\}\, ,
\end{eqnarray}
where $0=\lambda_1<\lambda_2<...<\lambda_n$ are the (real) eigenvalues of the Jacobian matrix $\bm{L}$. 
This first requirement is reasonable in the sense that a disturbance that varies too fast will not spread in the network, and thus either its source is rather easy to locate or it has no influence on the observed nodes. 

In order to guarantee that the system remains in the vicinity of the initial fixed point, we also require that the amplitude of $\xi_{\rm n,l}$ is not too large, 
\begin{eqnarray}\label{eq:ass_amp}
 \max_t|\xi_{\rm n}(t)| \ll \omega_i^0\, , \qquad
 \max_t|\xi_{\rm l}(t)| \ll a_{ij}^0\, .
\end{eqnarray}
These last assumptions allow us to consider that the system remains in the linear approximation regime. 
Again, this assumption is reasonable because a disturbance with too large amplitude would compromise the normal operation of the system, and its location then becomes a secondary problem. 
We will assume Eqs.~(\ref{eq:ass_freq}) and (\ref{eq:ass_amp}) to be satisfied throughout the manuscript.

\subsection{The Sherman-Morrison-Woodbury formula}
Our results rely heavily on the \emph{Sherman-Morrison-Woodbury formula}~\cite[Sec. 2.1.4]{golub2013matrix}, giving an explicit formulation for the inverse of the rank-$k$ perturbation of an invertible matrix $\bm A$,
\begin{eqnarray}\label{eq:smw}
 \left(\bm A + \bm U \bm V\right)^{-1} = \bm A^{-1} - \bm A^{-1}\bm U(\bm I + \bm V\bm A^{-1}\bm U)^{-1}\bm V\bm A^{-1}\, ,
\end{eqnarray}
where $\bm U\in \mathbb{R}^{n\times k}$ and $\bm V\in\mathbb{R}^{k\times n}$ characterize the rank-$k$ perturbation. 
We emphasize that the Sherman-Morrison-Woodbury formula applies to the pseudoinverse of Laplacian matrices (even though such matrices are singular), provided that the columns (resp. rows) of $\bm{U}$ (resp. $\bm{V}$) are orthogonal to the kernel of $\bm A$. 

In the case where $k=1$, Eq.~(\ref{eq:smw}) reduces to
\begin{eqnarray}\label{eq:sherman}
 \left(\bm A + \bm{u}\bm{v}^\top\right)^{-1} = \bm A^{-1} - \frac{\bm A^{-1}\bm{u}\bm{v}^\top \bm A^{-1}}{1 + \bm{v}^\top \bm A^{-1}\bm{u}}\, ,
\end{eqnarray}
where $\bm{u}$ and $\bm{v}$ are vectors characterizing the rank-1 perturbation. 
Eq.~(\ref{eq:sherman}) is usually referred to as the \emph{Sherman-Morrison formula}. 

\subsection{The Kron reduction}
In the context of electrical networks, it is possible to rewrite the equation governing the currents (the power flow equations~\cite{machowski2008power}) with respect to a restricted number of voltage variables through \emph{Kron reduction}~\cite{dorfler2013kron}. 
This is done by taking the Shur complement~\cite[Sec. 3.2.11]{golub2013matrix} of the coupling matrix with respect to a subset of nodes and adapting the velocities accordingly. 
Indeed, this reduction is not restricted to power grids and can be applied to the coupling matrix of any network. 
Partitioning the nodes in two sets $I_g=\{1,...,n_g\}$ (the nodes that are not reduced) and $I_c=\{n_g+1,...,n_g+n_c\}$ (the ones that are reduced), we write the elements of Eq.~(\ref{eq:lin_fix}) in block form (reordering indices if necessary) 
\begin{eqnarray}
 \bm{x} = \left(
 \begin{array}{c}
  \bm{x}^g \\ \bm{x}^c
 \end{array}\right)\, , 
 \quad
 \bm{\omega} = \left(
 \begin{array}{c}
  \bm{\omega}^g \\ \bm{\omega}^c
 \end{array}\right)\, , 
 \quad
 \bm{L} = \left(
 \begin{array}{cc}
  \bm{L}^{gg} & \bm{L}^{gc} \\
  \bm{L}^{cg} & \bm{L}^{cc}
 \end{array}\right)\, .
\end{eqnarray}

The Kron-reduced coupling matrix is then the Schur complement of $\bm{L}^{cc}$ in $\bm{L}$, 
\begin{eqnarray}\label{eq:Lr}
 \bm{L}^r = \bm{L}^{gg} - \bm{L}^{gc}(\bm{L}^{cc})^{-1}\bm{L}^{cg}\, ,
\end{eqnarray}
and applying a similar reduction to the vector of natural velocities gives 
\begin{eqnarray}\label{eq:wr}
 \bm{\omega}^r = \bm{\omega}^g - \bm{L}^{gc}(\bm{L}^{cc})^{-1}\bm{\omega}^c\, .
\end{eqnarray}
This yields the Kron-reduced version of Eq.~(\ref{eq:lin_fix}), which is restricted to the non-reduced nodes 
\begin{eqnarray}
 \bm{\omega}^r = \bm{L}^r\bm{x}^g\, ,
\end{eqnarray}
which allows to solve the equations for the subset of nodes $I_g$. 

Note that if the coupling matrix is Laplacian, then its Kron-reduced counterpart is also Laplacian.

\section{Locating a single disturbance}\label{sec:method}
We assume the disturbance to vary sufficiently slowly, meaning that its effect will propagate throughout the network. 
The detection of the disturbance can then be performed at any point of the system (e.g., through Fourier transform of the signal), but locating it requires a more elaborate procedure. 
We use the same method to identify disturbances at nodes or on lines. 
In order to perform this we need to know the Jacobian matrix $\bm{L}$ of the system, possibly Kron-reduced if some nodes are not measured, i.e., we need to be informed about the unperturbed state of the system. 
Under our assumption of relatively small amplitude in ${\bm \omega}$, a reasonable approximation of a stable fixed point of Eq.~(\ref{eq:dyn}) is 
\begin{eqnarray}\label{eq:angles}
 \bm{x}^g = (\bm{L}^{r})^\dagger\bm{\omega}_r\, ,
\end{eqnarray}
where $\dagger$ denotes the \emph{Moore-Penrose pseudoinverse} of the matrix~\cite[Sec. 5.5.2]{golub2013matrix}. 
Note that it then satisfies $\bm{L}\bm{L}^\dagger=\mathbb{I}-n^{-1}\bm{1}\bm{1}^\top$ because the Kron reduction $\bm{L}^r$ of $\bm{L}$ is a Laplacian matrix.  

The effect of the disturbances we consider translates into time-varying (reduced) Jacobian matrix $\tilde{\bm L}^r(t)$ and/or velocity vector $\tilde{\bm \omega}^r(t)$. 
As we assumed the amplitude of the disturbance to be relatively small, the evolution of the state of the system is reasonably approximated as $\bm{x}^g(t)\approx[\tilde{\bm L}^r(t)]^\dagger\tilde{\bm \omega}^r(t)$. 
If we have access to the time series of $\bm{x}^g(t)$ and the knowledge of $\bm{L}^r$, we can compute 
\begin{eqnarray}\label{eq:psi_def}
 \bm{\psi}(t) = \bm{L}^r\bm{x}^g(t) = \bm{L}^r\left[\tilde{\bm L}^r(t)\right]^\dagger\tilde{\bm \omega}^r(t)\, ,
\end{eqnarray}
which we refer to as the \emph{frequency mismatch}. 
We show now that the time series of $\bm{\psi}(t)$ are able to locate the disturbed element with as much accuracy as can be expected. 

We distinguish three cases: a disturbance at a node (which can be reduced or not); a line disturbance between two nodes that are not Kron-reduced; and a line disturbance with at least one end-node that is Kron-reduced.

\subsection{Nodal disturbance}\label{sec:nodal}
This case is the easiest to treat. 
The Jacobian matrix is constant in time $\tilde{\bm L}^r(t)=\bm{L}^r$ and only one component of the vector of velocities is time-varying. 
There are now two possible cases. 
If the perturbed node is not reduced, we can locate it exactly. 
Indeed, in this case, the reduced velocity vector has the form 
\begin{eqnarray}
 \tilde{\bm \omega}^r(t) = \tilde{\bm \omega}^g(t) - \bm{L}^{gc}(\bm{L}^{cc})^{-1}\bm{\omega}^c = \bm{\omega}^r + \xi_{\rm n}(t)\bm{e}_i\, ,
\end{eqnarray}
which, once plugged into Eq.~(\ref{eq:psi_def}), yields
\begin{eqnarray}
 \bm{\psi}(t) = \bm{\omega}^r + \xi_{\rm n}(t)\left(\bm{e}_i - n^{-1}\bm{1}\right)\, .
\end{eqnarray}
Provided that $n$ is not too small, the amplitude of $\bm{\psi}(t)$ will be significantly larger at the perturbed node, allowing then to identify it. 

If the perturbed node is Kron-reduced, the reduced velocity vector is expressed as 
\begin{eqnarray}
 \tilde{\bm \omega}^r(t) = \bm{\omega}^g - \bm{L}^{gc}(\bm{L}^{cc})^{-1}\tilde{\bm \omega}^c(t)\, .
\end{eqnarray}
Again, plugging this into Eq.~(\ref{eq:psi_def}), we see that all non-reduced nodes that are neighbors of the reduced component are significantly impacted by the disturbance. 
We can then identify the area of the network in which the faulty node is located, but we cannot identify the node exactly. 
This is not surprizing as we do not have a good enough accuracy in the measurements.

\subsection{Line disturbance between non-reduced end-nodes}
This is the most interesting case because we are able to locate the faulty line exactly. 
It also covers the case where no Kron reduction is applied. 
Here the faulty line has no influence on the reduced velocity vector and is a rank-1 perturbation of the reduced Jacobian matrix, 
\begin{eqnarray}\label{eq:lines1}
 \tilde{\bm \omega}^r(t) = \bm{\omega}^r\, , 
 \qquad 
 \tilde{\bm L}^r(t) = \bm{L}^r + \xi_{\rm l}(t)\bm{e}_{ij}\bm{e}_{ij}^\top\, ,
\end{eqnarray}
where we adapted the size of $\bm{e}_{ij}$ and the indices $i$ and $j$ in order to comply with the Kron reduction. 
Introducing Eq.~(\ref{eq:lines1}) into the computation of the frequency mismatch yields
\begin{eqnarray}
 \bm{\psi}(t) &= \bm{L}^r\left[\bm{L}^r + \xi_{\rm l}(t)\bm{e}_{ij}\bm{e}_{ij}^\top\right]^\dagger\bm{\omega}^r \nonumber \\
 &= \bm{L}^r\left[(\bm{L}^r)^\dagger - \frac{\xi_{\rm l}(t)(\bm{L}^r)^\dagger\bm{e}_{ij}\bm{e}^\top_{ij}(\bm{L}^r)^\dagger}{1 + \xi_{\rm l}(t)\bm{e}_{ij}^\top(\bm{L}^r)^\dagger\bm{e}_{ij}} \right]\bm{\omega}^r \nonumber \\ 
 &= \bm{\omega}^r - \alpha(t)\left[\bm{e}_{ij}^\top(\bm{L}^r)^\dagger\bm{\omega}^r\right]\bm{e}_{ij}\, , \label{eq:psi}
\end{eqnarray}
where we used Sherman-Morrison Eq.~(\ref{eq:sherman}) at the second line and defined
\begin{eqnarray}
 \alpha(t) = \frac{\xi_{\rm l}(t)}{1 + \xi_{\rm l}(t)\bm{e}_{ij}^\top(\bm{L}^r)^\dagger\bm{e}_{ij}}\, ,
\end{eqnarray} 
where we used that $\bm{L}\bm{L}^\dagger\bm{e}_{ij}=\bm{e}_{ij}$ which follows from the property of the pseudoinverse discussed below Eq.~(\ref{eq:angles})\,.
One sees that the only time varying components of $\bm{\psi}(t)$ are at the two end points of the perturbed line, which allows to locate it exactly. 

\begin{figure*}
 \centering
 \includegraphics[width=\textwidth]{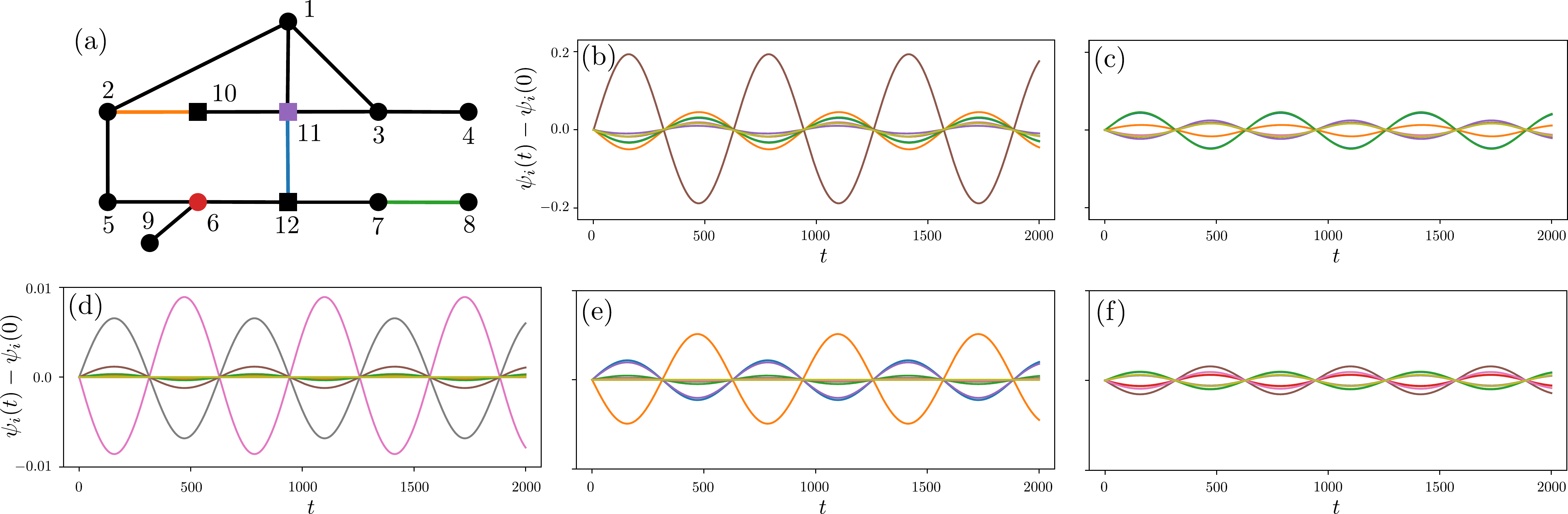}
 \caption{Example network with the five possible types and location of disturbances in Kron-reduced network.
 We consider an oscillating disturbance, whose frequency is much smaller than the time scales of the network. 
 (a) Network before Kron reduction. 
 The three square nodes $\{10,11,12\}$ are not measured and thus are reduced. 
 The colored nodes are subject to disturbances [panels (b) and (c)], as well as the three colored lines [panels (d), (e), and (f)]. 
 (b-f) Time series of $\bm{\psi}(t)$ as defined in Eq.~(\ref{eq:psi_def}). 
 The correspondence between indices and colors is given by $(1,2,3,4,5,6,7,8,9)\leftrightarrow({\rm blue, orange, green, red, purple, brown, pink, gray, yellow})$.
 (b) Disturbance at the non-reduced node $6$. 
 (c) Disturbance at the reduced node $11$. 
 (d) Disturbance on line $(7,8)$. 
 None of the end-nodes are reduced. 
 (e) Disturbance on line $(2,10)$ where node $10$ is reduced. 
 (f) Disturbance on line $(11,12)$, where both nodes are reduced. 
 }
 \label{fig:kron_impact}
\end{figure*}

This case is illustrated in Fig.~\ref{fig:kron_impact}(d), which shows the time series of $\psi_i(t)$ for nodes $1$ to $9$ of the network shown in Fig.~\ref{fig:kron_impact}(a), when the coupling of the green line is varying with time. 
One sees that the perturbed line is unambiguously identified. 

{\bf Remark.}
{\it We notice that the larger the weight of the disturbed line (in the Jacobian matrix ${\bm L}$), the more accurate the identification. 
Indeed, if the value $L_{ij}$ is small, its variation will hardly be noticed on the dynamics of the agents. 
This means that our method is the most reliable for disturbances occurring on the most important lines of the system. }

\subsection{Line disturbance with at least one reduced end-node}
In this situation, the exact location of the perturbed line cannot be determined based on the measurements because at least one of the end points is hidden to the observer. 
We detail the case where the two end points of the faulty line are in the Kron-reduced set. 
The mixed case where only one of the end points is reduced is very similar. 
All of our computations are inspired from~\cite[Secs. V.1 and V.2]{coletta2018performance} and rely on an appropriate use of the Sherman-Morrison formula Eq.~(\ref{eq:sherman}) and of the formulae for the Kron reduction Eqs.~(\ref{eq:Lr}) and (\ref{eq:wr}). 

Assuming that $i$ and $j$ are in the Kron-reduced set, one gets time-varying reduced velocity vector and time-varying reduced Jacobian matrix
\begin{eqnarray}\label{eq:cc1}
 \tilde{\bm \omega}^r(t) = \bm{\omega}^r - \beta(t) \left[\bm{e}_{ij}^\top(\bm{L}^{cc})^{-1}\bm{\omega}^c\right]\bm{v}\, , \\
 \tilde{\bm L}^r(t) = \bm{L}^r - \beta(t) \bm{v}\bm{v}^\top\, ,
\end{eqnarray}
where we defined
\begin{eqnarray}\label{eq:cc2}
 \beta(t) = \frac{\xi_{\rm l}(t)}{1 + \xi_{\rm l}(t)\bm{e}_{ij}^\top(\bm{L}^{cc})^{-1}\bm{e}_{ij}}\, ,
\end{eqnarray}
and
\begin{eqnarray}\label{eq:cc3}
 \bm{v} = \bm{L}^{gc}(\bm{L}^{cc})^{-1}\bm{e}_{ij}\, ,
\end{eqnarray}
$\bm{e}_{ij}$ is adapted according to the Kron reduction, and we used that $\bm{L}$ is symmetric, i.e., $\bm{L}^{gc}=(\bm{L}^{cg})^\top$. 
Now in the same spirit as before, the frequency mismatch is 
\begin{eqnarray}
 \bm{\psi}(t) &= \bm{L}^r[\tilde{\bm L}^r(t)]^\dagger\bm{\omega}^r(t) \nonumber \\
 &= \bm{L}^r\left[(\bm{L}^r)^\dagger - \frac{\beta(t)(\bm{L}^r)^\dagger\bm{v}\bm{v}^\top(\bm{L}^r)^\dagger}{1 + \beta(t)\bm{v}^\top(\bm{L}^r)^\dagger\bm{v}}\right] \nonumber \\
 &\phantom{=} \times \left[\bm{\omega}^r - \beta(t)\left(\bm{e}_{ij}^\top(\bm{L}^{cc})^{-1}\bm{\omega}^c\right)\bm{v}\right] \nonumber \\
 &= \bm{\omega}^r + \gamma(t)\bm{v}\, , \label{eq:cc4}
\end{eqnarray}
where again, we used the Sherman-Morrison formula, Eq.~(\ref{eq:sherman}), and where we gathered all time-varying quantities in $\gamma(t)$. 

For the case where node $i$ is not reduced and node $j$ is, a similar calculation gives 
\begin{eqnarray}
 \bm{\psi}(t) = \bm{\omega}^r + \tilde{\gamma}(t)\tilde{\bm v}\, ,
\end{eqnarray}
where
\begin{eqnarray}\label{eq:wt}
 \tilde{\bm v} = \bm{e}_i + \bm{L}^{gc}(\bm{L}^{cc})^{-1}\bm{e}_j\, ,
\end{eqnarray}
with $\bm{e}_i$ and $\bm{e}_j$ being adapted to the Kron-reduced system, and where, again, we assembled all time-varying quantities in $\tilde{\gamma}(t)$.

In these two cases, the effect of the perturbed line will be measured [through $\bm{\psi}(t)$] on the non-reduced nodes connected to the reduced component containing the perturbed line. 
This can be verified by inspection of the submatrix $\bm{L}^{gc}$ and is seen in Figs.~\ref{fig:kron_impact}(e) and (f), showing the time series of $\psi_i(t)$, when the disturbed line is at the blue and orange lines respectively. 
One cannot identify precisely its location, but it is possible to identify the reduced component to which it belongs, which is the most we could expect from the available measurements. 
In Fig.~\ref{fig:kron_impact}(e), where the perturbed line connects a reduced node to a non-reduced one, we observe that the amplitude of the frequency mismatch $\bm{\psi}(t)$ is much larger at the non-reduced end of the disturbed line than at any other non-reduced node. 
This makes sense if line couplings are all similar, the $i$th component of $\tilde{\bm v}$ [Eq.~(\ref{eq:wt})] is likely to be significantly larger than its other components, which will translate as a larger amplitude of variation in $\psi_i(t)$. 
However, in full generality, we cannot guarantee that the amplitude of $\psi_i(t)$ will be larger than all other components of the frequency mismatch. 

In summary, by simply computing the frequency mismatch, we are able to locate node and line disturbances, at locations in the system that are not necessarily measured.

\section{Multiple disturbances}\label{sec:multi_dist}
Interestingly, we observe that our approach allows to locate multiple disturbances acting simultaneously, provided they have sufficiently different characteristics so one is able to distinguish them. 
We treat here the case of two simultaneous disturbances, the case with more disturbances extends naturally from it. 
We show the cases where the disturbances occur in non-reduced areas. 
The reduced case works similarly as in the single disturbance case.

For two nodal disturbances $\xi_1(t)$ and $\xi_2(t)$, at non-reduced nodes $i$ and $j$ respectively, the calculation of the frequency mismatch yields,
\begin{eqnarray}
 \bm{\psi}(t) = \bm{\omega}^r + \xi_1(t)(\bm{e}_i - n^{-1}\bm{1}) + \xi_2(t)(\bm{e}_j - n^{-1}\bm{1})\, .
\end{eqnarray}
Provided the disturbances $\xi_1$ and $\xi_2$ have sufficiently different characteristics, one can identify their location if $n$ is not too small. 
If one of the nodes is reduced, only the area of the disturbance can be located. 

Two simultaneous line disturbances can be modelled as the following rank-2 perturbation of the Jacobian matrix,
\begin{eqnarray}
 \tilde{\bm L}(t) = \bm{L} + 
 \left(
 \begin{array}{cc}
  \bm{e}_{ij} & \bm{e}_{k\ell} 
 \end{array}
 \right)
 \cdot
 \left(
 \begin{array}{c}
    \xi_1(t)\bm{e}_{ij}^\top  \\  \xi_2(t)\bm{e}_{k\ell}^\top 
 \end{array}\right) = \bm{L} + \bm{U} \cdot \bm{V}_t\,, 
\end{eqnarray}
where lines $(i,j)$ and $(k,\ell)$ are perturbed. 
As in the single disturbance case, one can then calculate ${\bm \psi}(t)$, which yields, using the Sherman-Morrison-Woodbury formula Eq.~(\ref{eq:smw}), 
\begin{eqnarray}
 \bm{\psi}(t) &= \bm{L}[\tilde{\bm L}(t)]^\dagger\bm{\omega} = \bm{L}\left[\bm{L}^{\dagger} - \bm{L}^\dagger \bm U\left(\mathbb{I} + \bm V_t\bm{L}^\dagger \bm U\right)^{-1}\bm V_t\bm{L}^\dagger\right]\bm{\omega} \nonumber \\
 &= \bm{\omega} -\bm U \underbrace{\left(\mathbb{I} + \bm V_t\bm{L}^\dagger \bm U\right)^{-1}}_{= \bm{X}}\bm V_t\bm{\theta}^*\, .\label{eq:psi_multi}
\end{eqnarray}
Recall that $\bm{\omega}$, the columns of $\bm U$, and the row of $\bm V_t$ are orthogonal to the kernel of $\bm{L}$. 

Parsing Eq.~(\ref{eq:psi_multi}), we see that 
\begin{eqnarray}
 {\bm V_t}\bm{\theta}^* = \left(\begin{array}{c} \xi_1(t)(\theta_i^*-\theta_j^*) \\ \xi_2(t)(\theta_k^*-\theta_\ell^*) \end{array}\right)\, .
\end{eqnarray}
Furthermore, we see numerically that the matrix $\bm X$ is close to be diagonal. 
Namely, the off-diagonal elements of $\bm X$ are smaller (in absolute value) than its diagonal elements by at least one order of magnitude. 
This yields 
\begin{eqnarray}\label{eq:guess}
 \bm{\psi}(t) \approx \bm{\omega} - X_{11}\xi_1(t)(\theta_i^*-\theta_j^*)\bm{e}_{ij} - X_{22}\xi_2(t)(\theta_k^*-\theta_\ell^*)\bm{e}_{k\ell}\, ,
\end{eqnarray}
where $X_{pp}$ is the $p$th diagonal element of $\bm X$. 
The disturbances can then be located unequivocally, provided they have sufficiently different characteristics. 

\begin{figure}
 \center
 \includegraphics[width=\textwidth]{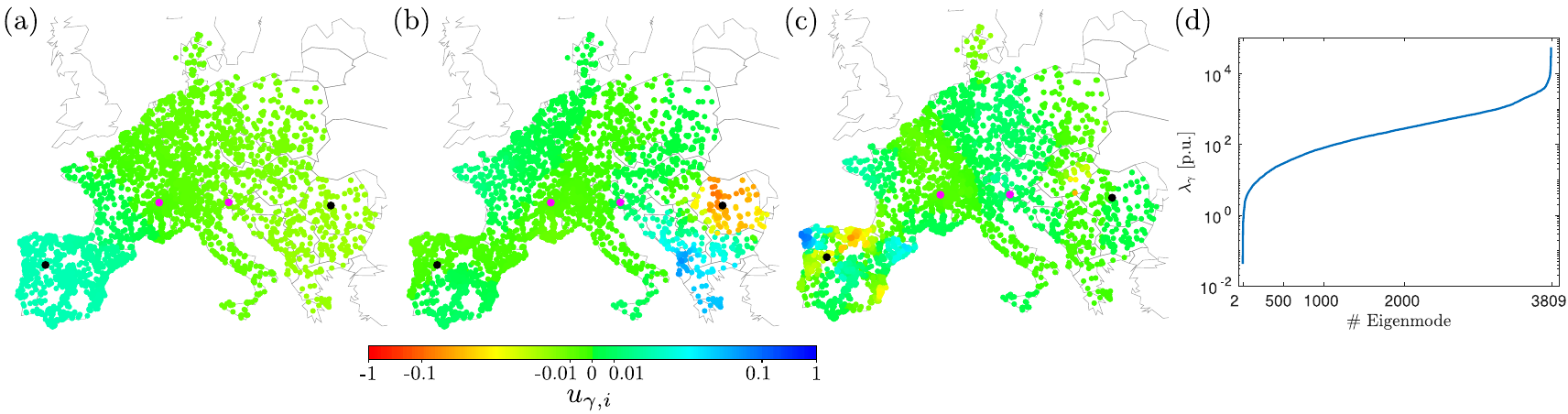}
 \caption{(a)-(c) 2nd, 10th and 20th eigenvectors of the European electric transmission grid. 
 (d) All of its eigenvalues, which vary over almost six order of magnitudes. 
 Purple and black dots display the locations of faulty lines in first and second columns of Fig.~\ref{fig:comparison} respectively.}
 \label{fig:mode}
\end{figure}

Even though we have not been able to certify analytically that $\bm X$ is almost diagonal, we explain it heuristically as follows. 
Analytically, everything boils down to showing that $\bm V_t\bm{L}^\dagger \bm U$ is (close to) diagonal. 
Let us forget the time dependence of $V$ for now ($\xi_1=\xi_2\equiv 1$) and compute the $(\alpha,\beta)$th term of this matrix, where we denote by $(i_\alpha,j_\alpha)$ the disturbed lines,
\begin{eqnarray}\label{eq:offd}
 (\bm V\bm{L}^\dagger \bm U)_{\alpha\beta} = \bm{e}_{i_\alpha j_\alpha}^\top\bm{L}^\dagger\bm{e}_{i_\beta j_\beta} = \sum_{\gamma\ge2} \frac{(u_{\gamma,i_\alpha}-u_{\gamma,j_\alpha})(u_{\gamma,i_\beta}-u_{\gamma,j_\beta})}{\lambda_\gamma} \, ,
\end{eqnarray}
where $\lambda_\gamma$ is the $\gamma$th eigenvalue of $\bm{L}$ and $u_{\gamma,k}$ denotes the $k$th component of the associated eigenvector. 
At first sight, one could think that the main contribution to Eq.~(\ref{eq:offd}) comes from eigenmodes with small eigenvalues, i.e., $\gamma$ close to $2$. However it appears that, for the corresponding eigenvectors, the components of neighboring nodes are rather similar (see Fig.~\ref{fig:mode}, showing some eigenmodes on a large-scale network modelling the European power transmission grid). 
It is particularly true in large-scale networks where these low laying eigenmodes spread over the whole network, at odds with high laying ones that are rather localized on a few nodes.  
One could think then, that the contribution comes from these high laying eigenmodes as the components of neighboring nodes might differ significantly~\cite{pagnier2019inertia}. 
However, these contributions also remain small due to the denominator in Eq.~(\ref{eq:offd}) where large values for $\lambda_\gamma$ bring down the contribution. 
The diagonal terms in $\bm{X}$ however are not vanishing due to the identity matrix appearing in Eq.~(\ref{eq:psi_multi}). 

We performed a numerical investigation of the matrix $\bm V\bm{L}^\dagger \bm U$ for the European grid. On average over all possible pairs of lines, the absolute value of its off-diagonal term is $5.2\cdot10^{-6}$  and it reaches $5.9\cdot10^{-2}$ for the worst configuration. This seems to validate that the assumption we made on $\bm X$ holds, at least for interaction graphs based on infrastructures.

Even though we have not been able to ground our observations in analytical results, it is remarkable that such a naive approach is able to isolate various disturbances in a complex system. 

\section{Implementation and numerical validation}\label{sec:nums}
In order to validate our approach, we simulated three instanciations of the dynamics Eq.~(\ref{eq:dyn}) with three different types of line disturbances, on three different networks. 

\subsection{Algorithm}
\begin{figure}
 \centering
 \includegraphics[width=.85\textwidth]{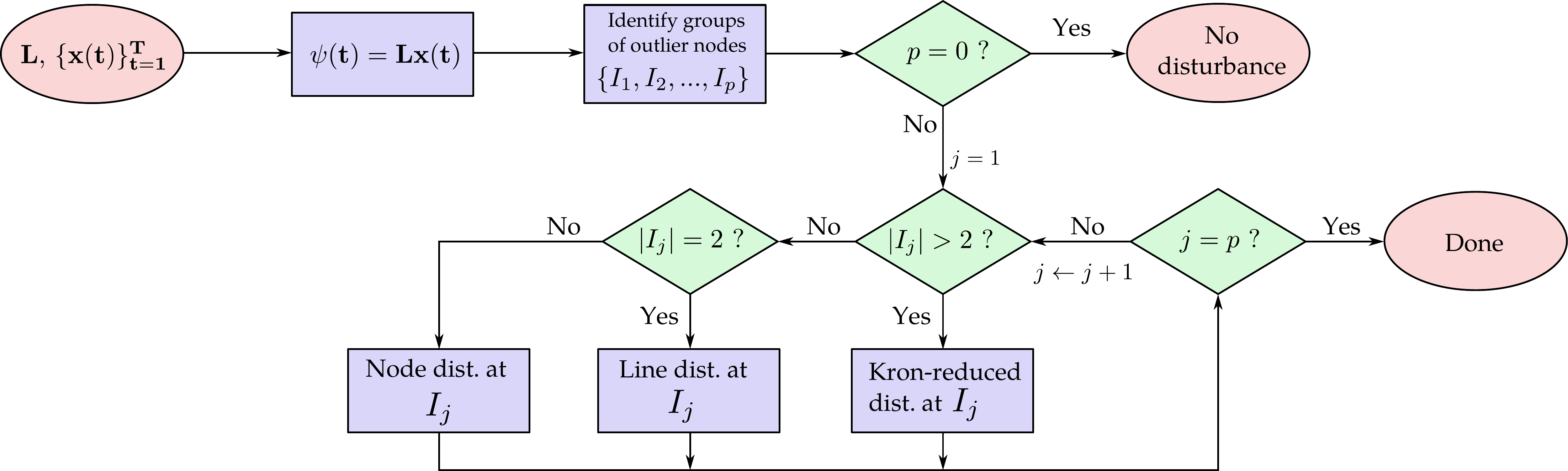}
 \caption{Block diagram of a suggested algorithm implementing our method. 
 Note that this manuscript covers the first block only, whose task is to ease the work of downstream blocks. 
 This does not mean that these subsequent blocks are trivial. }
 \label{fig:flowchart}
\end{figure}

We propose here and summarize in Fig.~\ref{fig:flowchart} the course of an algorithm that would implement our method. 

First, with the Jacobian matrix $\bm{L}$ and the times series $\{\bm{x}(t)\}_{t=1}^T$ as input, one can straightforwardly compute the frequency mismatch $\{\bm{\psi}(t)\}_{t=1}^T$. 
Then, the algorithm needs to identify groups of outlier nodes, whose frequency mismatch trajectory are significantly different from the bulk of all measured nodes. 
On top of that, these outlier nodes need to be grouped in sets ($I_1,...,I_p$) of qualitatively similar trajectories, e.g., blue and green (resp. orange and red) curves are grouped together in Figs.~\ref{fig:euroroad_noise} and \ref{fig:usairports_ramp_step}.
Defining what "significantly different" means is not always trivial and might be problem dependent, but it is necessary in order to identify disturbed nodes. 
Furthermore, identifying "significantly different" curves and classifying them by similarity might by a challenging problem for data analysts, but these issues are beyond the scope of our expertise in general and of this manuscript in particular. 

Once the nodes have been classified, each subset of nodes should correspond to a disturbance in the system. 
In Fig.~\ref{fig:flowchart}, we propose a way to determine the type of disturbance associated to each subset $I_j$. 

\subsection{PanTaGruEl}
The PanTaGruEl model of the interconnected European grid~\cite{pagnier2019inertia,tyloo2019key} is built on publicly available data of geolocalization of power system elements, and parameters (admittances, productions, loads,...) are reconstructed based on standard assumptions. 
It consists of 3809 nodes and 7343 lines. 
For our simulations, we consider an oscillating disturbance, 
\begin{eqnarray}\label{eq:oscil}
 \xi_{\rm l}(t) = \xi_0\sin(\omega_{\rm m} t)\, , 
\end{eqnarray}
satisfying the conditions detailed in Sec.~\ref{ssec:dist}. 
From a practical point of view, this allows us to tune the time scale of the disturbance, which is the oscillation frequency $\omega_{\rm m}$ in order to guarantee that it is sufficiently slow with respect to all time scales of the network. 

\begin{figure*}
 \center
 \includegraphics[width=\textwidth]{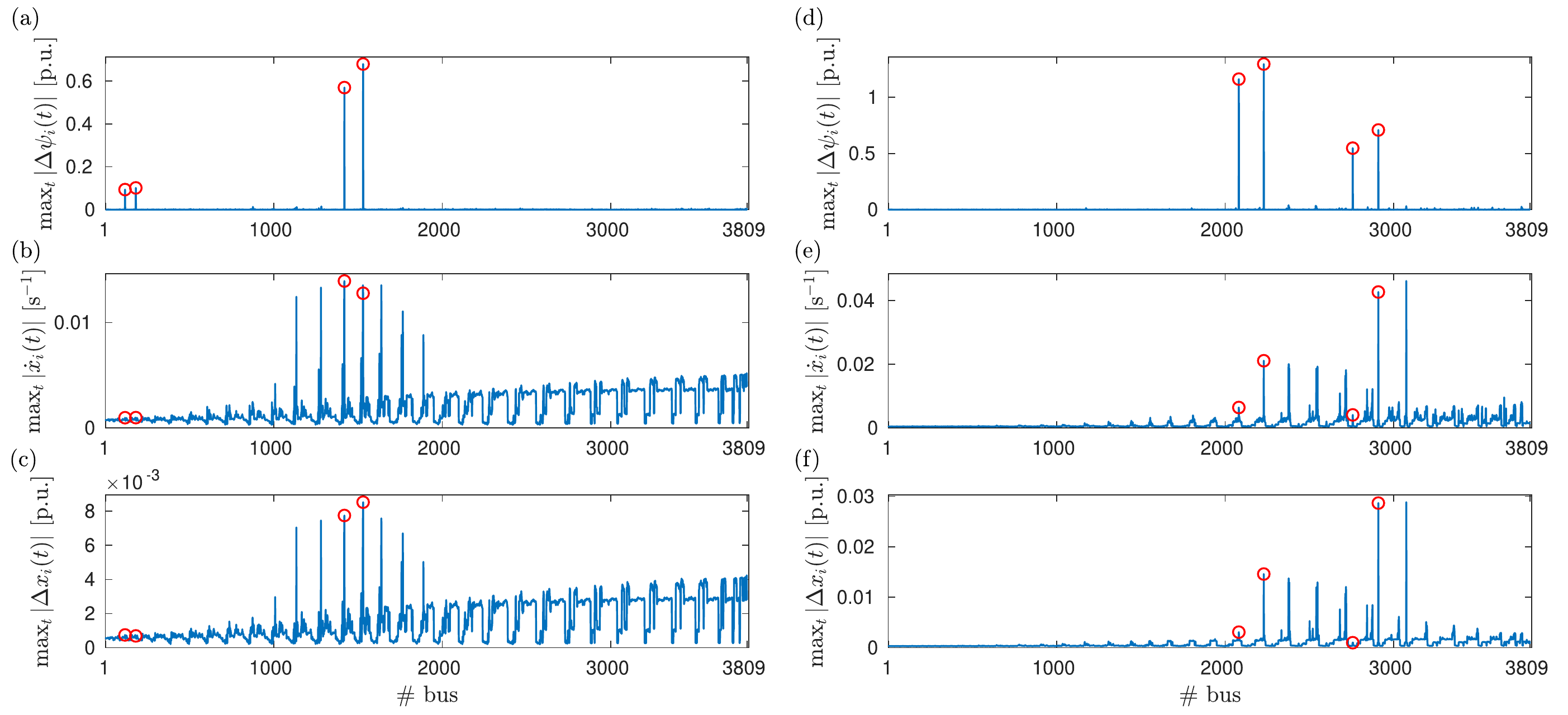}
 \caption{Comparison of the detection of perturbed lines based on $\psi_i(t)$, $\dot{x}_i(t)$ or $x_i(t)$, in the PanTaGruEl, a model of the European electric transmission grid. 
 The two columns show two configurations with different malfunctioning elements. In each panel, the end nodes of the two disturbed lines are indicated by the red circles are their locations are indicated in Fig.~\ref{fig:mode}.
 Reasonable noise was added at each node in order to challenge the robustness of our approach. 
 The disturbed lines are exactly identified as the ones whose end nodes have a significant amplitude for $\psi_i$. }
 \label{fig:comparison}
\end{figure*}

In the following, we compare the disturbance detection using our method relying on $\bm{\psi}(t)$ or by simply measuring the agents' frequencies $\dot{\bm x}(t)$ and value $\bm{x}(t)$.
The left panels of Fig.~\ref{fig:comparison} shows the maximal amplitude of $\psi_i(t)$ for the oscillating disturbance of five different lines of the PanTaGruEl model, shown in Fig.~\ref{fig:mode}. 
Lines are perturbed one at a time, in different simulations.  
The right panels show the maximal amplitude of the frequencies $\dot{x}_i(t)$ for the same time series. 
The red circle indicate the indices of the two extremities of the disturbed line. 
Our method identifies unequivocally the endpoints of the disturbed lines whereas the time series of frequencies are not able to do so.

In the case of sparse (incomplete) measurements, we have tested, but not shown in this works, our method for faults where both ends of the perturbed line belong to the reduced (non-measured) nodes. 
In this case, our methods correctly detect the nodes in the vicinity of the faulty elements. 
However, at this task, our method is not significantly better than the detection proceeding by inspection of the trajectories of $x_i(t)$\,, $\dot{x}_i(t)$\,.

\subsection{Euroroad and US Airports}
\begin{figure*}
 \centering
 \includegraphics[width=\textwidth]{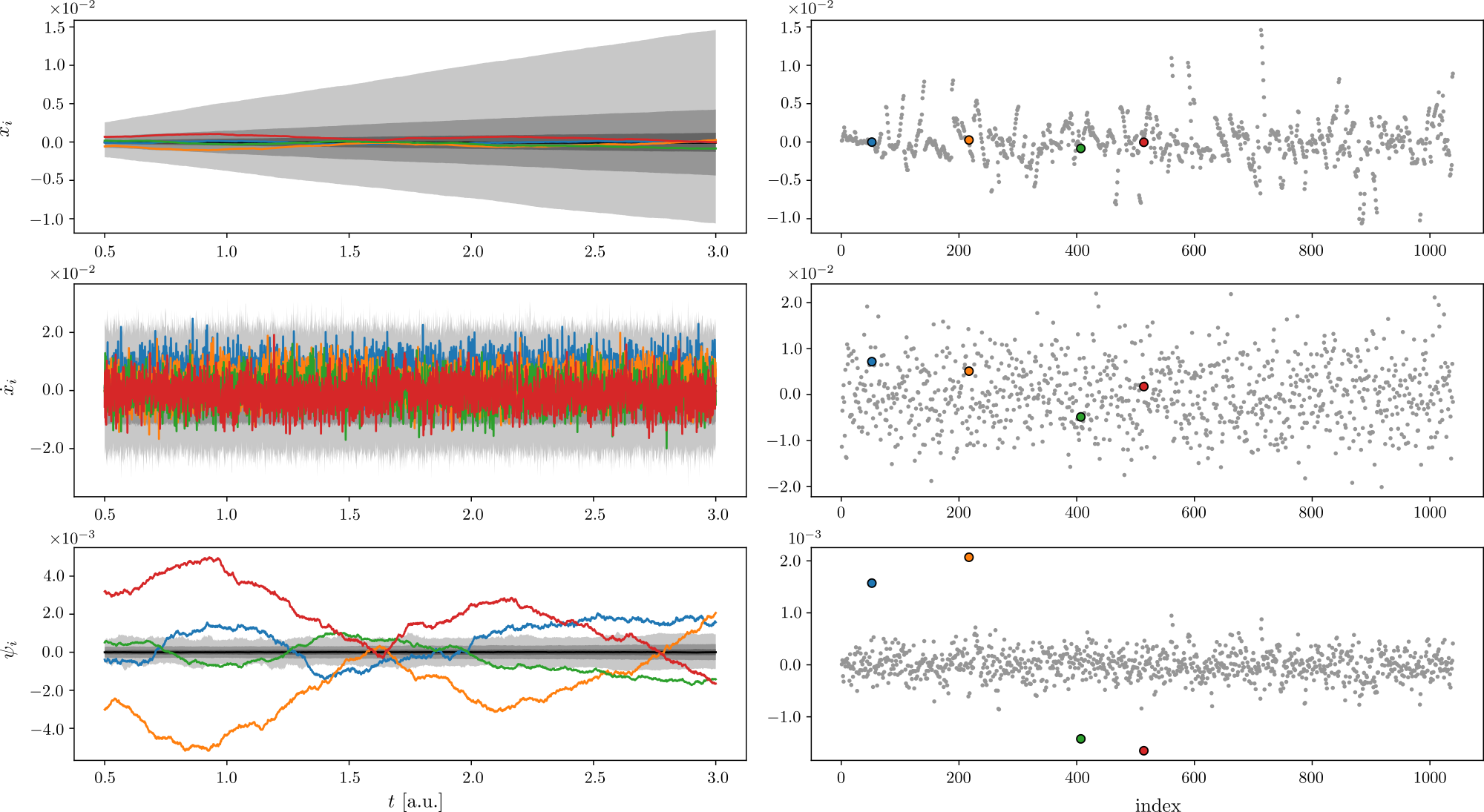}
 \caption{Phases, velocities, and frequency mismatch for the European road network~\cite{subelj2011robust,peixoto2020the} with linear coupling and two noisy lines. 
 The pairs of trajectories blue-green and orange-red correspond to the end nodes of each disturbed lines. 
 The grey area covers the same quantities for all other nodes. 
 The disturbances cannot be seen in the phases and velocities trajectories, whereas they are clearly identified and located with the frequency mismatch. 
 The panels on the right side are the snapshot at the end of the time series of the left side. }
 \label{fig:euroroad_noise}
\end{figure*}

\begin{figure*}
 \centering
 \includegraphics[width=\textwidth]{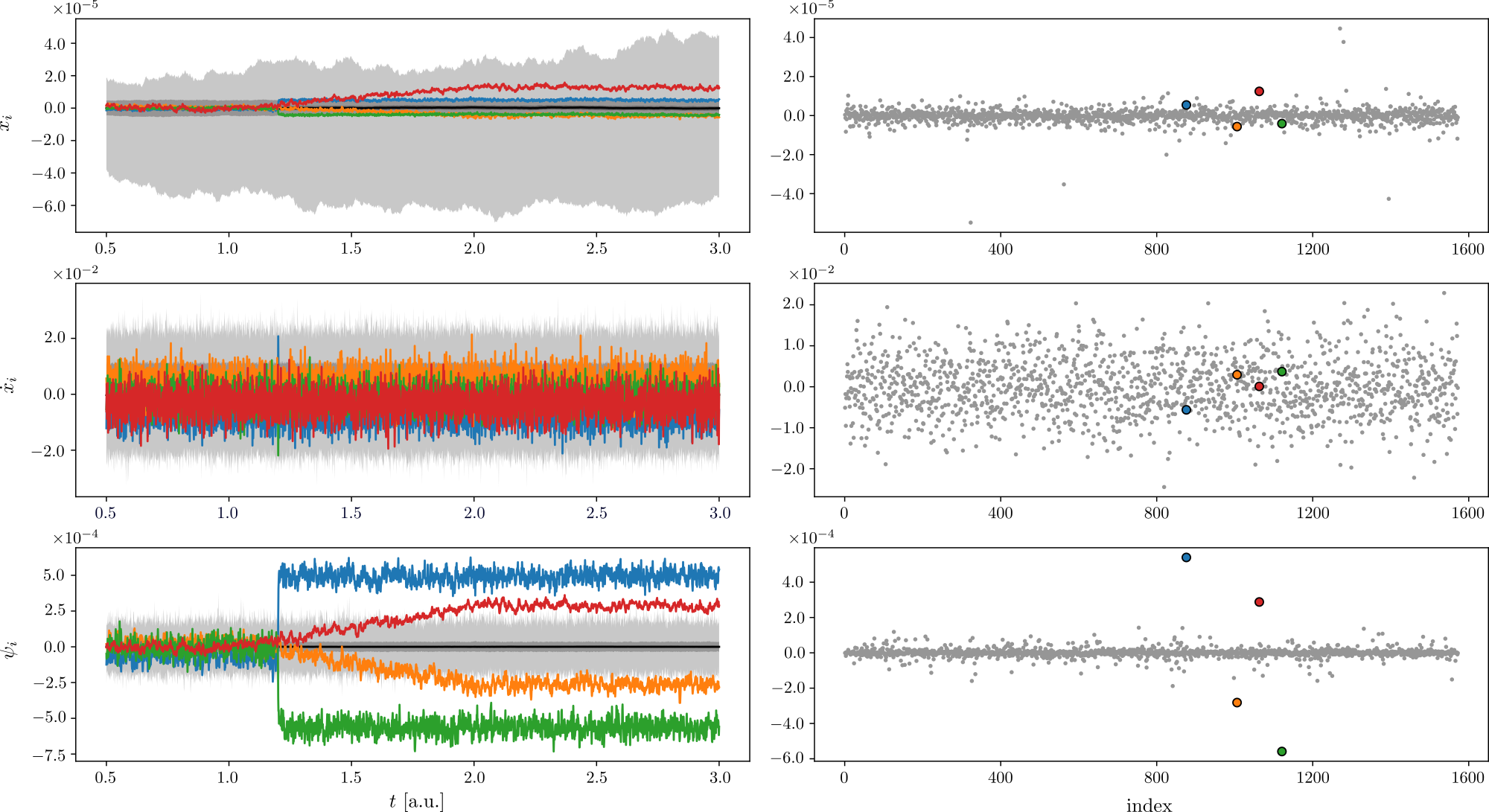}
 \caption{
 Phases, velocities, and frequency mismatch for the US airports network~\cite{guimera2005worldwide,colizza2007reactiondiffusion} with third order Kuramoto coupling~\cite{skardal2011cluster}. 
 One line is disturbed by a step signal and the other by a ramp signal. 
 The pairs of trajectories blue-green and orange-red correspond to the end nodes of each disturbed lines. 
 The grey area covers the same quantities for all other nodes. 
 The disturbances cannot be seen in the phases and velocities trajectories, whereas they are clearly identified and located with the frequency mismatch. 
 The panels on the right side are the snapshot at the end of the time series of the left side. 
}
 \label{fig:usairports_ramp_step}
\end{figure*}

In order to illustrate the variety of couplings that our method covers, we applied it to:
\begin{itemize}
 \item A set of $n=1039$ agents linearly coupled according to the largest component of the European road network from~\cite{subelj2011robust,peixoto2020the}. 
 Two arbitrary lines are pertubed by a noisy signal. 
 Results are shown in Fig.~\ref{fig:euroroad_noise};
 \item A set of $n=1572$ agents interacting through a \emph{higher-order Kuramoto coupling}~\cite{skardal2011cluster} (couplings of order up to $q=3$ are used here), with interaction structure given by the largest connected component of the network of US airports from~\cite{guimera2005worldwide,colizza2007reactiondiffusion}. 
 One line is perturbed by a ramp signal and another one is subject to a step signal. 
 Results are shown in Fig.~\ref{fig:usairports_ramp_step}. 
\end{itemize}
In each case, we added a reasonable amount of noise at each node, for sake of generality. 

In both Figs.~\ref{fig:euroroad_noise} and \ref{fig:usairports_ramp_step}, the left panels show the time series of $x_i(t)$, $\dot{x}_i(t)$, and $\psi_i(t)$ from top to bottom. 
The right panels show the snapshot of the same quantities at the end of the time series on the left, at $t=3.0$. 

We acknowledge that, even though it is straighforward to identify the end-nodes of the disturbed lines by visual inspection of Figs.~\ref{fig:euroroad_noise} and \ref{fig:usairports_ramp_step}, it might be an algorithmic challenge to automatize this process. 

\section{Conclusion}\label{sec:conclusion}
We proposed an elegant method to identify and locate disturbances in networked dynamical systems. 
Our method relies on time series of the agents degree of freedom and is able to differentiate between line disturbances and nodal disturbances. 
In the case of partial measurements, we are able to locate precisely the perturbation if the nodes where the it occurs are measured. 
Otherwise, we can determine the area of the network where the faulty, unmeasured nodes/lines are located, using the Kron-reduced network. 
The main condition required for our method to work out well is that the disturbance changes much slower than the intrinsic time scales of the system. However it is not restricted to a particular shape for the perturbation or a specific dynamical model, making it rather widely applicable. Interestingly, our method is also able to locate multiple disturbance occurring at the same time, given that they have different characteristics. This represents a remarkable feature as in large scale networks made of thousands or millions of nodes, many disturbances are likely to overlap in the time series. 

We believe that the idea raised in this manuscript will help the development of more efficient tools to detect and locate disturbances more accurately. 
Future work will aim at proposing a scheme for online detection and localization of disturbances.

\section*{Acknowledgments}
RD and MT were supported by the Swiss National Science Foundation under grant no. 200020\_182050. 
RD was supported by ETH Zurich funding and by the Swiss National Science Foundation under grant no. P400P2\_194359.

\appendix

\section{Directed networks}\label{sec:directed}
Although all mathematical objects involved in the derivations of Sec.~\ref{sec:method} are well-defined for directed graphs, the results are harder to desintricate. 
With directed interactions, the nodal perturbation is similar to the undirected case, 
\begin{eqnarray}
 \tilde{\bm \omega}(t) = \bm{\omega} + \xi_{\rm n}(t)\bm{e}_i\, ,
\end{eqnarray}
and the line disturbance has the form 
\begin{eqnarray}
 \tilde{\bm L}(t) = \bm{L} + \xi_{\rm l}(t)\bm{e}_i\bm{e}_{ij}\, .
\end{eqnarray}

In any case, the computation $\bm{\psi}(t)$ involves a vector of the form 
\begin{eqnarray}
 \bm{w} = \bm{L}\bm{L}^\dagger \bm{e}_i\, ,
\end{eqnarray}
which is the othogonal projection of $\bm{e}_i$ on the image of $\bm{L}$~\cite[Sec.~5.5.2]{golub2013matrix}. 
One can verify that $\bm{w}$ highly depends on the position of node $i$ in the network and the structure thereof. 
Desintricating the effect of the network structure on the vector $\bm{w}$ is beyond the scope of this manuscript and we defer it to future work.

\end{document}